\documentclass[a4paper,12pt]{article}

\makeatletter
 
  \@addtoreset{equation}{section}
 \makeatother

\usepackage{amsfonts}
\usepackage{amssymb}
\usepackage{mathrsfs}
\usepackage{amsmath,amsthm}
\usepackage{amsmath}

\usepackage{graphicx}
\usepackage{color}
\usepackage{indentfirst}

\makeatletter
\def\@biblabel#1{}
\makeatother

\usepackage{cite}

\begin{document}

\newtheorem{theorem}{Theorem}
\newtheorem{lemma}{Lemma}
\newtheorem{proposition}{Proposition}
\theoremstyle{definition}
\newtheorem{defn}{Definition}
\newtheorem{remark}{Remark}
\newtheorem{step}{Step}
\newtheorem{cor}{Corollary}

\newcommand{\Cov}{\mathop {\rm Cov}}
\newcommand{\Var}{\mathop {\rm Var}}
\newcommand{\E}{\mathop {\rm E}}
\newcommand{\const }{\mathop {\rm const }}
\everymath {\displaystyle}

\newcommand{\ruby}[2]{
\leavevmode
\setbox0=\hbox{#1}
\setbox1=\hbox{\tiny #2}
\ifdim\wd0>\wd1 \dimen0=\wd0 \else \dimen0=\wd1 \fi
\hbox{
\kanjiskip=0pt plus 2fil
\xkanjiskip=0pt plus 2fil
\vbox{
\hbox to \dimen0{
\small \hfil#2\hfil}
\nointerlineskip
\hbox to \dimen0{\mathstrut\hfil#1\hfil}}}}

\def\qedsymbol{$\blacksquare$}
\renewcommand{\thefootnote }{\fnsymbol{footnote}}

\renewcommand{\refname }{References}

\title{Theoretical Sensitivity Analysis for Quantitative Operational Risk Management
\footnote{Forthcoming in {\it International Journal of Theoretical and Applied Finance}.}}
\author{Takashi Kato
\footnote{Association of Mathematical Finance Laboratory (AMFiL), 
              2--10, Kojimachi, Chiyoda, Tokyo 102-0083, Japan, 
E-mail: \texttt{takashi.kato@mathfi-lab.com}}
}
\date{First Version: April 3, 2011\\
This Version: May 24, 2017}
\maketitle

\begin{abstract}
We study the asymptotic behavior of the difference between 
the values at risk $\mathrm {VaR}_\alpha (L)$ and $\mathrm {VaR}_\alpha (L+S)$ for 
heavy-tailed random variables $L$ and $S$ with $\alpha \uparrow 1$ 
for application in sensitivity analysis of quantitative operational risk management 
within the framework of the advanced measurement approach of Basel II (and III). 
Here $L$ describes the loss amount of the present risk profile 
and $S$ describes the loss amount caused by an additional loss factor. 
We obtain different types of results 
according to the relative magnitudes of the thicknesses of the tails of $L$ and $S$. 
In particular, if the tail of $S$ is sufficiently thinner than that of $L$, then 
the difference between prior and posterior risk amounts 
$\mathrm {VaR}_\alpha (L+S) - \mathrm {VaR}_\alpha (L)$ is asymptotically equivalent to 
the expectation (expected loss) of $S$. 
\end{abstract}

\everymath {\displaystyle}

\section{Introduction}\label{sec_intro}

Value at risk (VaR) is a standard risk measure and is widely used in quantitative financial risk management. 
In this paper, we study the asymptotic behavior of VaRs for heavy-tailed random variables (loss variables). 
We focus on the asymptotic behavior of the difference between two VaRs of 
\begin{eqnarray}\label{diff_VaR}
\Delta \mathrm {VaR}_\alpha ^{L,S} := \mathrm {VaR}_\alpha (L+S) - \mathrm {VaR}_\alpha (L)
\end{eqnarray}
with $\alpha \rightarrow 1$, where $L$ and $S$ are regularly varying random variables and 
$\alpha \in (0, 1)$ denotes the confidence level of VaR. 
For our main result, we show that the behavior of $\Delta \mathrm {VaR}_\alpha ^{L,S}$ 
($\alpha \rightarrow 1$) drastically changes according to the relative magnitudes of the thicknesses of the tails of $L$ and $S$. 
Interestingly, when the tail of $S$ is sufficiently thinner than that of $L$, 
$\Delta \mathrm {VaR}_\alpha ^{L,S}$ is approximated by the expected loss (EL) of $S$, that is, 
$\Delta \mathrm {VaR}^{L, S}_\alpha \sim \E [S]$, $\alpha \rightarrow 1$. 

This study is strongly related to quantitative operational risk management within the framework of the Basel Accords, 
which is a set of recommendations for regulations in the banking industry. 
Here, we briefly introduce the financial background.

Basel II (International Convergence of Capital Measurement and Capital Standards: A Revised Framework) 
was published in 2004, and in it, operational risk was added as a new risk category
(see Basel Committee on Banking Supervision (2004) (BCBS) 
and 
\cite {Embrechts-et-al2} 
for the definition of operational risk). 
Beginning in 2013, the next accord, Basel III, was scheduled to be introduced in stages. 
Note that the quantitative method for measuring operational risk was not substantially changed in Basel III at this stage,  
so here we refer to mainly Basel II. 

To measure the capital charge for operational risk, banks may choose among three approaches: 
the basic indicator approach, the standardized approach, and an advanced measurement approach. 
While the basic indicator approach and the standardized approach prescribe explicit formulas, 
the advanced measurement approach does not specify a model to quantify a risk amount (risk capital), and hence banks adopting this approach must construct their own quantitative risk model and 
verify it periodically
\footnote{In 2016, BCBS is proposing to remove the advanced measurement approach from the regulatory framework (Pillar I) in the finalization of Basel III (also known as Basel IV). 
Needless to say, this examination does not mean that advanced methods for risk measurement are unnecessary. 
It is still important in practice to utilize internal models to improve risk profiles. 
See \cite {Basel_SMA1}, \cite {Basel_SMA2}, and \cite {PwC}.}. 

\cite {Embrechts-et-al2} pointed out that
although everyone agrees on the importance of understanding operational risk, 
controversy remains regarding how far one should (or can) quantify such risks. 
Because empirical studies have found 
that the distribution of operational loss has a fat tail 
(see \cite {Moscadelli}), 
we should focus on capturing the tail of the loss distribution. 

Basel II does not specify a measure of the risk but states that 
``a bank must be able to demonstrate that its approach captures potentially severe `tail' loss events. 
Whatever approach is used, a bank must demonstrate that its operational risk measure meets 
a soundness standard comparable to that of the internal ratings-based approach for credit risk, 
(\textit{i.e.}, comparable to a one-year holding period and a $99.9^{\mbox {th}}$ percentile confidence interval)'' 
in paragraph 667 of 
\cite {BaselII}. 
A typical risk measure, which we adopt in this paper, is VaR at the confidence level $0.999$. 
Meanwhile, estimating the tail of an operational loss distribution is 
often difficult for a number of reasons, such as insufficient historical data and the various kinds of factors in operational loss. 
We therefore need sufficient verification of the appropriateness and robustness of the model used
in quantitative operational risk management. 

One verification approach for a risk model is sensitivity analysis (also called model behavior analysis). 
The term ``sensitivity analysis'' can be interpreted in a few different ways. 
In this paper, we use this term to mean analysis of the relevance of a change in the risk amount with changing 
input information (\textit{e.g.}, added/deleted loss data or changing model parameters). 
Using sensitivity analysis is advantageous 
not only for validating the accuracy of a risk model but also for deciding
the most effective policy with regard to the variable factors. 
This examination of how the variation in the output of a model can be apportioned to different sources of 
variations in risk will give impetus to business improvements. 
Moreover, sensitivity analysis is also meaningful for scenario analysis. 
Basel II requires not only use of historical internal/external data
and Business Environment and Internal Control Factors (BEICFs) as input information, 
but also use of scenario analyses 
to evaluate low-frequency high-severity loss events that cannot be captured by empirical data. 
As noted above, to quantify operational risk we need to estimate the tail of 
the loss distribution, so it is important to recognize 
the impact of our scenarios on the risk amount. 

In this study, we perform a sensitivity analysis of the operational risk model from a theoretical viewpoint. 
In particular, we consider mainly the case of adding loss factors. 
Let $L$ be a random variable that represents the loss amount with respect to the present risk profile 
and let $S$ be a random variable of the loss amount caused by an additional loss factor 
found by a detailed investigation or brought about by expanded business operations. 
In a practical sensitivity analysis, it is also important to consider the statistical effect 
(estimation error of parameters, \textit{etc.}) when validating an actual risk model, but 
such an effect should be treated separately. 
We focus on the change from a prior risk amount $\rho (L)$ to a posterior risk amount $\rho (L+S)$, 
where $\rho $ is a risk measure. 
We use VaR at the confidence level $\alpha $ as our risk measure $\rho $ and 
we study the asymptotic behavior of VaR as $\alpha \rightarrow 1$. 

Our framework is mathematically related to the study of 
\cite {Bocker-Kluppelberg2}. 
They regard $L$ and $S$ as loss amount variables of separate categories (cells) 
and study the asymptotic behavior of an aggregated loss amount $\mathrm {VaR}_\alpha (L+S)$ as $\alpha \rightarrow 1$. 
In addition, a similar study that adopts a conditional VaR (CVaR; also called an expected shortfall, average VaR, tail VaR, etc.) 
is found in 
\cite {Biagini-Ulmer}. 

In other related work, 
\cite {Degen-et-al2} study the asymptotic behavior of the ``risk concentration'' 
\begin{eqnarray}\label{risk_concentration}
C_\alpha ^{L, S} = \frac{\mathrm {VaR}_\alpha (L+S)}{\mathrm {VaR}_\alpha (L) + \mathrm {VaR}_\alpha (S)}
\end{eqnarray} 
with $\alpha \rightarrow 1$ to investigate the risk diversification benefit when $L$ and $S$ are identically distributed. 
Similarly, \cite {Embrechts-et-al3} study (\ref {risk_concentration}) for multivariate regularly varying random vector $(L, S)$ to examine asymptotic super-/sub-additivity of VaRs. 
The properties of (\ref {risk_concentration}) are also studied in \cite {Embrechts-Neslehova-Wuthrich}, \cite {Embrechts-Puccetti-Ruschendorf}, and \cite {Embrechts-Wang-Wang} with respect to risk aggregation and model robustness. 

In contrast, in this work, we aim to estimate the difference in VaRs, such as (\ref {diff_VaR}), so our results give higher order estimates than the above results. 
In addition, our result does not require the assumption that $L$ and $S$ are identically distributed. 
Moreover, we do not specify a model for $L$ and $S$, so our results are applicable to the general case. 

Our results also have the following financial implications. 
\begin{itemize}
 \item [(a)] When $S$ is a loss amount with high frequency and low severity, the effect of adding $S$ is estimated by capturing the EL of $S$; thus, we should focus on the body of the distribution of $S$ rather than the tail, 
 \item [(b)] When $S$ is a loss amount with low frequency and high severity, we should take care to capture the tail of $S$. 
\end{itemize}
Similar empirical and statistical arguments are found in \cite {Frachot-et-al}, and 
we provide the theoretical rationale for these arguments. 

The rest of this paper is organized as follows. 
In Section \ref {sec_model}, we introduce the framework of our model and some notation. 
In Section \ref {sec_sum_VaR}, we give rough estimations of the asymptotic behavior of the risk amount $\mathrm {VaR}_\alpha (L+S)$. 
Our main results are in Section \ref {sec_indep}, and we present finer estimations of the difference between $\mathrm {VaR}_\alpha (L)$ and $\mathrm {VaR}_\alpha (L+S)$, 
that is, $\Delta \mathrm {VaR}_\alpha ^{L, S}$. 
In Section \ref {sec_ex}, we present some numerical examples of our results. 
Section \ref {sec_conclusion} concludes the paper. 
Section \ref {sec_appendix} is an appendix. 
In Section 
\ref {sec_generalization}, 
we give a partial generalization of the results in Section \ref {sec_indep} 
when $L$ and $S$ are not independent. 
One of these results is related to the study of risk capital decomposition and 
we study these relations in Section 
\ref {sec_EL2}. 
All the proofs of our results are given in Section 
\ref {sec_proofs}.

\section{Notation and Model Settings}\label{sec_model}

We fix the probability space as $(\Omega, \mathcal {F}, P)$, and assume that all random variables below are defined on it.
For a random variable $X$ and $\alpha \in (0, 1)$, we define the $\alpha $-quantile (VaR) by 
\begin{eqnarray}
\mathrm {VaR}_\alpha (X) = \inf \{ x\in \Bbb {R}\ ; \ F_X(x) \geq \alpha  \} , 
\end{eqnarray} 
where $F_X(x) = P(X\leq x)$ is the distribution function of $X$. 

We denote by $\mathcal {R}_k$ the set of regularly varying functions with index $k\in \Bbb {R}$; 
that is, $f\in \mathcal {R}_k$ if and only if $\lim _{x\rightarrow \infty }f(tx)/f(x) = t^k$ 
for any $t > 0$. When $k = 0$, a function $f\in \mathcal {R}_0$ is called slowly varying. 
For the details of regular variation and slow variation, see 
\cite{Bingham-et-al} and 
\cite {Embrechts-et-al}. 
For a random variable $X$, we denote $X\in \mathcal {R}_k$ when 
the tail probability function $\bar{F}_X(x) = 1 - F_X(x) = P(X > x)$ is in $\mathcal {R}_k$. 
We mainly treat the case of $k < 0$, in which the $m$th moment of $X\in \mathcal {R}_k$ is infinite for $m > -k$. 
Two well-known examples of heavy-tailed distributions that have regularly varying tails are
the generalized Pareto distribution (GPD) and the g-h distribution 
(see 
\cite {Degen-et-al}, 
\cite {Dutta-Perry}), which are widely used in quantitative operational risk management. 
In particular, the GPD plays an important role in extreme value theory, 
and can approximate 
the excess distributions over a high threshold of all the commonly used continuous distributions. 
See 
\cite {Embrechts-et-al} and 
\cite {Embrechts-et-al2} 
for details. 

Let $L$ and $S$ be non-negative random variables and assume 
$L\in \mathcal {R}_{-\beta }$ and $S \in \mathcal {R}_{-\gamma }$ for some $\beta , \gamma > 0$. 
We call $\beta $ (resp. $\gamma $) the tail index of $L$ (resp. $S$). 
A tail index represents the thickness of a tail probability. 
For example, the relation $\beta < \gamma $ means that the tail of $L$ is fatter than that of $S$. 

We regard $L$ as the total loss amount of a present risk profile. 
In the framework of the standard loss distribution approach 
(LDA, see 
\cite {Frachot-et-al} 
for details), 
$L$ is often assumed to follow a compound Poisson distribution. 
If we consider a multivariate model, $L$ is given by $L = \sum ^d_{k=1}L_k$, 
where $L_k$ is the loss amount variable of the $k$th operational risk cell ($k = 1, \ldots , d$). 
We are aware of such formulations, but we do not limit ourselves to such situations in our settings. 

The random variable $S$ means an additional loss amount. 
We consider the total loss amount variable $L + S$ as a new risk profile. 
As mentioned in Section \ref {sec_intro}, our interest is in how a prior risk amount $\mathrm {VaR}_\alpha (L)$ changes to 
a posterior amount $\mathrm {VaR}_\alpha (L + S)$. 

In Sections \ref {sec_sum_VaR}--\ref {sec_ex}, 
we assume that $L$ and $S$ are always independent. 
The assumption of independence implies that 
the loss events are caused independently by factor $L$ or $S$. 
The case where the loss events are not independent is studied in Section \ref {sec_generalization}. 

\section{Basic Results of Asymptotic Behavior of $\mathrm {VaR}_\alpha (L + S)$}\label{sec_sum_VaR}

First we give a rough estimation of $\mathrm {VaR}_\alpha (L + S)$ 
in preparation for introducing our main results in the next section. 

\begin{proposition} \ \label{prop_sum_VaR} \ 
\begin{description}
 \item[ \mbox{(i)} ] \ If $\beta < \gamma $, then $\mathrm {VaR}_\alpha (L + S) \sim \mathrm {VaR}_\alpha (L)$, 
 \item[ \mbox{(ii)} ] \ If $\beta = \gamma $, then $\mathrm {VaR}_\alpha (L + S) \sim \mathrm {VaR}_{1-(1-\alpha )/2}(U)$, 
 \item[ \mbox{(iii)} ] \ If $\beta > \gamma $, then $\mathrm {VaR}_\alpha (L + S) \sim \mathrm {VaR}_\alpha (S)$ 
\end{description} 
as $\alpha \rightarrow 1$, where the notation $f(x)\sim g(x)$, \ $x\rightarrow a$ denotes 
$\lim _{x\rightarrow a}f(x) / g(x) = 1$ and $U$ is a random variable 
whose distribution function is given by $F_U(x) = (F_L(x) + F_S(x)) / 2$. 
\end{proposition}

These results are easily obtained and not novel. 
In particular, 
this proposition is 
strongly related to the results in 
\cite{Bocker-Kluppelberg2} 
(in the framework of LDA).

In contrast with Theorem 3.12 in 
\cite{Bocker-Kluppelberg2}, 
which implies an estimate for 
$\mathrm {VaR}_\alpha (L+S)$ as ``an aggregation of $L$ and $S$'', 
we review the implications of Proposition \ref {prop_sum_VaR} from the viewpoint of sensitivity analysis. 
Proposition \ref {prop_sum_VaR} implies that 
when $\alpha $ is close to one, 
the posterior risk amount is determined nearly entirely by either of the risk amounts 
$L$ or $S$ showing a fatter tail. 
On the other hand, when the thicknesses of the tails are the same (\textit{i.e.}, $\beta = \gamma $,) 
the posterior risk amount $\mathrm {VaR}_\alpha (L + S)$ is given by the VaR of the random variable $U$ 
and is influenced by both $L$ and $S$ even if $\alpha $ is close to one. 

\begin{remark}
The random variable $U$ is a variable determined by a fair coin toss (heads: $L$;  tails: $S$). 
That is, $U$ is defined as $U = L1_{\{B = 0\}} + S1_{\{B = 1\}}$, where 
$B$ is a random variable with the Bernoulli distribution (\textit{i.e.}, $P(B = 0) = P(B = 1) = 1/2$) which is independent of $L$ and $S$. 
Then it holds that 
$P(U\leq x) = (P(L\leq x) + P(S\leq x)) / 2$. 
Therefore, $F_U$ defined in Proposition \ref {prop_sum_VaR} is actually a distribution function. 
\end{remark}

\begin{remark}
Note that we can generalize the results of Proposition \ref {prop_sum_VaR} to the case 
where $L$ and $S$ are not independent. 
For instance, we can easily show that assertions (i) and (iii) always hold in general. 
One of the sufficient conditions for assertion (ii) is a ``negligible joint tail condition''; 
that is, 
\begin{description}
 \item[ \mbox{[A1]}] \ 
$P(L > x, S > x)/(\bar{F}_L(x) + \bar{F}_S(x)) \longrightarrow 0$ as 
$x\rightarrow \infty $
\end{description}
(see 
\cite {Jang-Jho}
for details). 
Related results are also obtained in 
\cite {Albrecher-et-al} and \cite {Geluk-Tang}. 
\end{remark}

\section{Main Results}
\label{sec_indep}

In this section, we present a finer representation of the difference between 
$\mathrm {VaR}_\alpha (L + S)$ and $\mathrm {VaR}_\alpha (L)$ than Proposition \ref {prop_sum_VaR}. 
We set the following conditions. 
\begin{description}
 \item[ \mbox{[A2]}] \ There is some $x_0 > 0$ such that 
$F_L$ has a positive, non-increasing density function $f_L$ on $[x_0, \infty )$; that is,
$F_L(x) = F_L(x_0) + \int ^x_{x_0}f_L(y)dy, \ \ x \geq x_0$. 
 \item[ \mbox{[A3]}] \ The function $x^{\gamma - \beta }\bar{F}_S(x)/\bar{F}_L(x)$ converges to 
some real number $k$ as $x\rightarrow \infty $. 
 \item[ \mbox{[A4]}] \ The same assertion as [A2] holds when $L$ is replaced with $S$. 
\end{description}

Note that condition [A2] (resp. [A4]) and the monotone density theorem 
(Theorem 1.7.2 in 
\cite {Bingham-et-al}) 
imply $f_L\in \mathcal {R}_{-\beta - 1}$ (resp. $f_S\in \mathcal {R}_{-\gamma  - 1}$).

\begin{remark}\label{rem_1}
The condition [A3] seems a little strict: it implies that 
$\mathcal {L}_L$ and (a constant multiple of) $\mathcal {L}_S$ are asymptotically equivalent, where 
$\mathcal {L}_L(x) = x^{\beta }\bar{F}_L(x)$ and $\mathcal {L}_S(x) = x^{\gamma }\bar{F}_S(x)$ are 
slowly varying functions. 
However, since the Pickands--Balkema--de Haan theorem 
(see 
\cite {Balkema-deHaan} and 
\cite {Pickands}) 
implies 
that $\bar{F}_L$ and $\bar{F}_S$ are approximated by the GPD, 
the asymptotic equivalence of $\mathcal {L}_L$ and $\mathcal {L}_S$ 
approximately holds 
(note that [A3] is guaranteed when $L$ and $S$ follow the GPD: see Section \ref {sec_ex}). 
Note that when $\beta = \gamma $, 
condition [A3] is called a ``tail-equivalence'' (see 
\cite {Embrechts-et-al} for instance). 
\end{remark}

Our main result is the following. 

\begin{theorem} \ \label{th_main}
Define $\Delta \mathrm {VaR}_\alpha ^{L, S}$ by (\ref {diff_VaR}) and 
similarly $\Delta \mathrm {VaR}_\alpha ^{S, L}$ with switching the roles of $L$ and $S$. 
Then the following assertions hold as $\alpha \rightarrow 1$. 
${\rm (i)}$  If $\beta + 1 < \gamma $, then 
$\Delta \mathrm {VaR}_\alpha ^{L, S}\sim \E [S]$ under $\mathrm {[A2]}$. 
${\rm (ii)}$  If $\beta < \gamma \leq \beta + 1$, then 
$\Delta \mathrm {VaR}_\alpha ^{L, S} \sim 
\frac{k}{\beta }\mathrm {VaR}_\alpha (L)^{\beta + 1 - \gamma }$ under $\mathrm {[A2]}$ and $\mathrm {[A3]}$. 
${\rm (iii)}$  If $\beta = \gamma $, then 
$\Delta \mathrm {VaR}_\alpha ^{L, S} \sim 
\{ (1 + k)^{1/\beta } - 1\} \mathrm {VaR}_\alpha (L)$ 
under $\mathrm {[A3]}$. 
${\rm (iv)}$  If $\gamma < \beta \leq \gamma + 1$, then 
$\Delta \mathrm {VaR}_\alpha ^{S, L} \sim  
\frac{1}{k\gamma }\mathrm {VaR}_\alpha (S)^{\gamma + 1 - \beta }$ 
under $\mathrm {[A3]}$ and $\mathrm {[A4]}$. 
${\rm (v)}$  If $\gamma + 1< \beta $, then 
$\Delta \mathrm {VaR}_\alpha ^{S, L} \sim  \E [L]$ 
under $\mathrm {[A3]}$ and $\mathrm {[A4]}$. 

\end{theorem}

The assertions of Theorem \ref {th_main} are divided into five cases according to relative magnitudes of $\beta $ and $\gamma $. 
In particular, when $\beta < \gamma $, we get different results depending on whether or not $\gamma $ is greater than $\beta + 1$. 
Assertion (i) implies that 
if the tail probability of $S$ is sufficiently lower than that  of $L$, 
then the effect of a supplement of $S$ is limited to the EL of $S$.
In fact, we also get a result similar to assertion (i), 
introduced in Section \ref {sec_EL2}, when the impact of $S$ is absolutely small. 
These results indicate that if an additional loss amount $S$ is not so large, 
we may not need to be concerned about the effect of a tail event raised by $S$. 

Assertion (ii) implies that when $\gamma \leq \beta + 1$, 
the difference in risk amount cannot be approximated by the EL even if $\gamma > 1$, 
and we need information on the tail (rather than the body) of the distribution of $S$. 
For instance, let us consider the case where $S$ describes scenario data $(l, p)\in (0, \infty )\times (0, 1)$ 
such that $P(S > l) = p$ and 
$l$ is large enough (or, equivalently, $p$ is small enough) that 
$\mathrm {VaR}_{1-p}(L) \geq \mathrm {VaR}_{1-p}(S) = l$ and $p \leq 1 - \alpha $. 
Then, we can formally interpret assertion (ii) as 
\begin{eqnarray}\nonumber 
\Delta \mathrm {VaR}_\alpha ^{L, S} &\sim & 
\frac{k}{\beta }\mathrm {VaR}_\alpha (L)^{\beta - \gamma }\mathrm {VaR}_\alpha (L) \leq 
\frac{k}{\beta }\mathrm {VaR}_{1-p}(L)^{\beta - \gamma }\mathrm {VaR}_\alpha (L)
\\\label{eq_linear2}
&\approx & 
\frac{1}{\beta }\left( \frac{l}{\mathrm {VaR}_{1-p}(L)}\right) ^\gamma \mathrm {VaR}_\alpha (L)
\leq  
\frac{1}{\beta }\left( \frac{l}{\mathrm {VaR}_{1-p}(L)}\right) ^\beta  \mathrm {VaR}_\alpha (L). 
\end{eqnarray}
Strictly speaking, if we also assume that $\lim _{x\rightarrow \infty }\mathcal {L}_S(x)$ exists and is finite ($\mathcal {L}_S$ is given in Remark \ref {rem_1}), 
we have $\bar {F}_L(x) \sim k^{-1}x^{\gamma - \beta }\bar {F}_S(x) 
\sim k^{-1}\bar {F}_S(x^{\beta / \gamma })$, $x\rightarrow \infty $. 
Applying Lemma \ref {lemma2} (with $\lambda = k^{-1}$), 
we get $\mathrm {VaR}_{1-p}(L)^{\beta - \gamma }\sim k^{-1}(\mathrm {VaR}_{1-p}(S)/\mathrm {VaR}_{1-p}(L))^\gamma $ for small $p$, hence (\ref {eq_linear2}) is verified. 
Thus, it is sufficient to provide the amount on the right-hand side of (\ref {eq_linear2}) 
for an additional risk capital. So, in this case, the information on the pair $(l, p)$ 
and detailed information on the tail of $L$ enable conservative estimation of the difference. 

When the tail of $S$ has the same thickness as the tail of $L$, we have assertion (iii). 
In this case, we see that by a supplement of $S$, 
the risk amount is multiplied by $(1 + k)^{1/\beta }$. 
The slower the decay speed of $\bar{F}_S(x)$, 
which means the fatter the tail amount variable becomes with an additional loss, 
the larger the multiplier $(1 + k)^{1/\beta }$. 
Moreover, if $k$ is small, we have the approximation
\begin{eqnarray}\label{eq_linear}
\Delta \mathrm {VaR}_\alpha ^{L, S}
\sim 
\left( \frac{k}{\beta } + o(k)\right) \mathrm {VaR}_\alpha (L). 
\end{eqnarray}
Here, $o(\cdot )$ is the Landau symbol (little o): $\lim _{k\rightarrow 0}o(k)/k = 0$. 
The relation (\ref {eq_linear}) has the same form as assertion (ii), and 
in this case we have a similar implication as (\ref {eq_linear2}) 
by letting $\alpha = 1 - p$ and $k = (l/\mathrm {VaR}_{1-p}(L))^\beta $. 

Assertions (iv)--(v) are merely restated consequences 
of assertions (i)--(ii). 
In these cases, $\mathrm {VaR}_\alpha (L)$ is too small compared with 
$\mathrm {VaR}_\alpha (L + S)$ and $\mathrm {VaR}_\alpha (S)$, 
so we need to compare $\mathrm {VaR}_\alpha (L + S)$ with 
$\mathrm {VaR}_\alpha (S)$. 
In estimating the posterior risk amount, $\mathrm {VaR}_\alpha (L+S)$, 
the effect of the tail index $\gamma $ of $S$ is significant. 

By Theorem \ref {th_main}, we see that 
the smaller the tail index $\gamma $, 
the more precise is the information needed about the tail of $S$. 

\begin{remark}It should be noted that 
assertion (iii) of Theorem \ref {th_main} is not novel. 
For instance, a more general result than (iii) is introduced in 
\cite {Barbe-et-al}
(without strict arguments: 
\cite {Barbe-et-al} have not provided a complete proof, 
so we give lemmas that immediately show Theorem \ref {th_main}(iii) in Section \ref {sec_proofs}). 
Moreover, assertions (iv)--(v) are mathematically the same as (i)--(ii) 
(thus, the mathematical contribution of our results is centered on the first two assertions). 
Nevertheless, we have stated assertions (iii)--(v) in Theorem \ref {th_main} 
for demonstrating how the asymptotic behavior of the difference in VaR changes 
by shifting the value of $\gamma $ (or $\beta $) as mentioned above. 
\end{remark}

\section{Numerical Examples}\label{sec_ex}

In this section we numerically verify our main results 
for typical examples in the standard LDA framework. 
Let $L$ and $S$ be given by the compound Poisson variables 
$L = L^1 + \cdots + L^N$, \ $S = S^1 + \cdots + S^{\tilde{N}}$, 
where $(L^i)_i, (S^i)_i, N, \tilde{N}$ are independent random variables and 
the elements of $(L^i)_i $, $(S^i)_i$ are each independent and identically distributed. 
The variables $N$ and $\tilde{N}$ mean the frequency of loss events, 
and the variables $(L^i)_i$ and $(S^i)_i$ indicate the severity of each loss event. 
We assume that $N \sim \mathrm {Poi}(\lambda _L)$ and $\tilde{N} \sim \mathrm {Poi}(\lambda _S)$ for some 
$\lambda _L, \lambda _S > 0$, where $\mathrm {Poi}(\lambda )$ denotes the Poisson distribution 
with intensity $\lambda $. 
To be strict, we use the GPD, whose distribution function is given by 
$\mathrm {GPD}(\xi ,\sigma )(x) = 1 - \left( 1 + \xi x / \sigma \right) ^{-1/\xi }, \ x \geq 0$. 

Throughout this section, we assume that $L^i$ follows $\mathrm {GPD}(\xi _L, \sigma _L)$ 
with $\xi _L= 2, \sigma _L= 10000$ and set $\lambda _L = 10$. 
We also assume that $S^i$ follows $\mathrm {GPD}(\xi _S, \sigma _S)$ and $\lambda _S = 10$. 
We set the values of parameters $\xi _S$ and $\sigma _S$ in each case appropriately. 
We note that $L\in \mathcal {R}_{-\beta }$ and $S\in \mathcal {R}_{-\gamma }$, 
where $\beta = 1/\xi _L$ and $\gamma = 1/\xi _S$. 
Moreover, condition [A3] is satisfied with 
\begin{eqnarray}\label{def_k}
k = \frac{\lambda _S}{\lambda _L}(\sigma _S / \xi _S)^{1/\xi _S}(\sigma _L / \xi _L)^{-1/\xi _L}. 
\end{eqnarray}
Indeed, we observe that 
\begin{eqnarray}
\frac{x^{\gamma - \beta }\bar {F}_{S_1}(x)}{\bar{F}_{L_1}(x)} = 
\frac{(1/x + \xi _L/\sigma _L)^{1/\xi _L}}{(1/x + \xi _S/\sigma _S)^{1/\xi _S}} \longrightarrow 
(\sigma _S / \xi _S)^{1/\xi _S}(\sigma _L / \xi _L)^{-1/\xi _L}, \ \ x\rightarrow \infty , \ \ \ \ \ 
\end{eqnarray} 
and Theorem 1.3.9 in \cite {Embrechts-et-al} tells us that 
$\bar {F}_L(x) \sim \lambda _L\bar {F}_{L_1}(x)$ and $\bar {F}_S(x) \sim \lambda _S\bar {F}_{S_1}(x)$ with $x\rightarrow \infty $. 

To calculate VaR in the framework of LDA, several numerical methods are known. 
Monte Carlo, Panjer recursion, and 
inverse Fourier (or Laplace) transform approaches are widely used (see 
\cite {Frachot-et-al}). 
The direct numerical integration of 
\cite {Luo-Shevchenko} 
is an adaptive method
for calculating VaR precisely when $\alpha $ is close to one, 
and a similar direct integration method has been proposed by 
\cite {Kato_FMA}. 
These approaches are classified 
as inverse Fourier transform approaches. 
The precision of these numerical methods was compared in 
\cite {Shevchenko}. 
We need to have quite accurate calculations, so we use direct numerical integration 
to calculate 
$\mathrm {VaR}_\alpha (L)$ and $\mathrm {VaR}_\alpha (L + S)$. 

Unless otherwise noted, we set $\alpha = 0.999$. 
Then, the value of the prior risk amount $\mathrm {VaR}_\alpha (L)$ is $5.01\times 10^{11}$.

\subsection{The case of $\beta + 1 < \gamma $}\label{ex_sec_case1_1}

First we consider the case of Theorem \ref {th_main}(i). 
We set $\sigma _S = 10000$. 
The result is given in Table \ref {table_case1_1}, where 
\begin{eqnarray}\label{def_error}
\mathrm {Error} = \frac{\mathrm {Approx}}{\Delta \mathrm {VaR}} - 1 
\end{eqnarray}
and $\mathrm {Approx} = \E [S]$. 
Here, $\Delta \mathrm {VaR}_\alpha ^{L, S}$ is simply denoted as $\Delta \mathrm {VaR}$. 

Although the absolute value of the error becomes slightly larger when $\gamma - \beta $ is near one, 
the difference in VaR is accurately approximated by $\E [S]$.

\subsection{The case of $\beta < \gamma \leq \beta + 1$}\label{ex_sec_case1_2}

\begin{figure}[!h]
\centerline{\includegraphics[height = 6cm,width=12cm, bb=10 566 587 835]{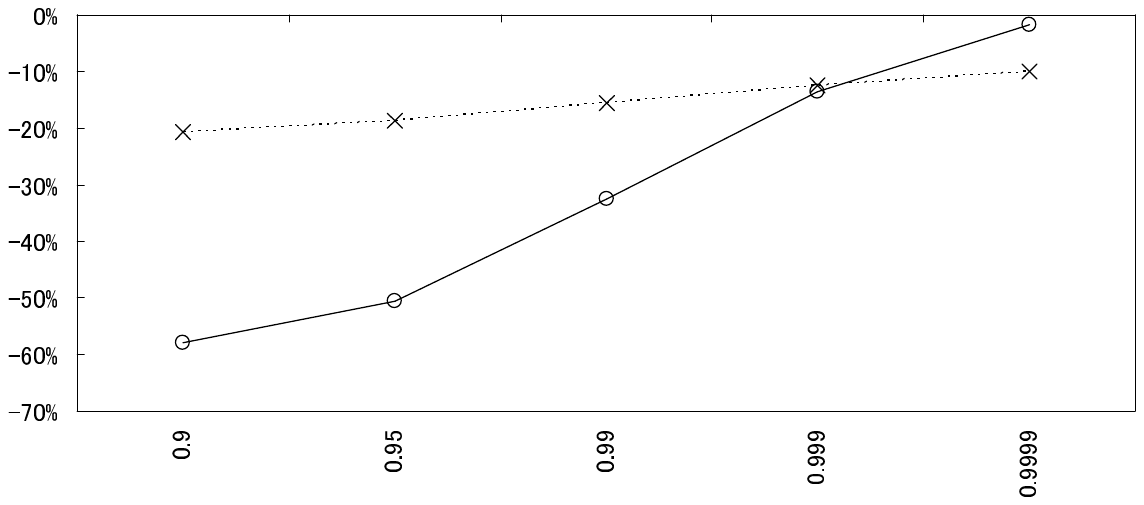}}
\caption{Change in approximation error via $\alpha $ in the cases of $\xi _S = 0.8$ 
(circles on solid line) and $\xi _S = 1.8$ (crosses on dotted line). 
The vertical axis corresponds to $\mathrm {Error}\times 100\% $ and the horizontal axis corresponds to $\alpha $. }
\label{fig_case1_3}
\end{figure}

This case corresponds to Theorem \ref {th_main}(ii). 
As in Section \ref {ex_sec_case1_1}, we also set $\sigma _S = 10,000$. 
The result is given as Table \ref {table_case1_2}, where 
$\mathrm {Approx} = k\mathrm {VaR}_\alpha (L)^{\beta + 1 -\gamma }/\beta $ and 
the error is the same as (\ref {def_error}). 
We see that the accuracy becomes lower when $\gamma - \beta $ is close to one or zero. 
Even in these cases, as Fig. \ref {fig_case1_3} shows for the cases of $\xi _S = 0.8$ and $\xi _S = 1.8$, 
we observe that the error approaches 0 when letting $\alpha \rightarrow 1$.

\begin{table}[!h]
\begin{center}
\caption{
Approximation errors in the case $\beta + 1 < \gamma $. 
$\xi _S$ is the shape parameter of the GPD with respect to $S$. 
$\beta $ (resp., $\gamma $) is the tail index of $L$ (resp., $S$). 
$\Delta \mathrm {VaR}$ is the same as in (\ref {diff_VaR}). 
Approx $= \E [S]$. 
Error is given by (\ref {def_error}). 
}
\label{table_case1_1}
\begin{tabular}{@{}ccccc@{}} \hline
 $\xi _S$	& $\gamma - \beta $	& $\Delta \mathrm {VaR}$	& Approx	& Error	\\ \hline
 0.1	& 9.500 	& 1,111,092	& 1,111,111	& 1.68E-05	\\ 
 0.2	& 4.500 	& 1,249,995	& 1,250,000	& 4.26E-06	\\ 
 0.3	& 2.833 	& 1,428,553	& 1,428,571	& 1.26E-05	\\ 
 0.4	& 2.000 	& 1,666,647	& 1,666,667	& 1.21E-05	\\ 
 0.5	& 1.500 	& 2,000,141	& 2,000,000	& -7.05E-05	\\ \hline
\end{tabular}
\end{center}
\end{table}

\begin{table}[!h]
\begin{center}
\caption{
Approximation errors in the case $\beta < \gamma \leq \beta + 1$. 
$\xi _S$ is the shape parameter of the GPD with respect to $S$. 
$\beta $ (resp., $\gamma $) is the tail index of $L$ (resp., $S$). 
$\Delta \mathrm {VaR}$ is the same as in (\ref {diff_VaR}). 
Approx $= k\mathrm {VaR}_\alpha (L)^{\beta + 1 -\gamma }/\beta $. 
Error is given by (\ref {def_error}). 
}
\label{table_case1_2}
\begin{tabular}{@{}ccccc@{}} \hline
 $\xi _S$	& $\gamma - \beta $	& $\Delta \mathrm {VaR}$	& Approx	& Error	\\ \hline
 0.8	& 0.750 	& 3.64E+06	& 3.14E+06	& -1.36E-01	\\ 
 1.0	& 0.500 	& 2.02E+08	& 2.00E+08	& -8.38E-03	\\ 
 1.2	& 0.333 	& 3.31E+09	& 3.30E+09	& -1.73E-03	\\ 
 1.5	& 0.167 	& 5.69E+10	& 5.63E+10	& -9.98E-03	\\ 
 1.8	& 0.056 	& 4.36E+11	& 3.81E+11	& -1.26E-01	\\ \hline
\end{tabular}
\end{center}
\end{table}

\subsection{The case of $\beta = \gamma $}\label{ex_sec_case2}

In this section, we set $\xi _S = 2 ( = \xi _L)$. We apply Theorem \ref {th_main}(iii). 
We compare the values of $\Delta \mathrm {VaR}$ and 
$\mathrm {Approx} = \{ (1+k)^{\xi _L}-1 \} \mathrm {VaR}_\alpha (L)$ in Table \ref {table_case2}, 
where the error is the same as (\ref {def_error}). 
We see that they are very close. 

\begin{table}[!h]
\begin{center}
\caption{
Approximation errors in the case $\beta = \gamma $. 
$\sigma _S$ is the scale parameter of the GPD with respect to $S$. 
$\Delta \mathrm {VaR}$ is the same as in (\ref {diff_VaR}). 
Approx $= \{ (1+k)^{\xi _L}-1 \} \mathrm {VaR}_\alpha (L)$. 
Error is given by (\ref {def_error}). 
}
\label{table_case2}
\begin{tabular}{@{}cccc@{}} \hline
 $\sigma _S$	& $\Delta \mathrm {VaR}$	& Approx	& Error	\\ \hline
 100	& 1.05E+11	& 1.05E+11	& -2.05E-07	\\ 
 1,000	& 3.67E+11	& 3.67E+11	& -1.85E-07	\\ 
 10,000	& 1.50E+12	& 1.50E+12	& -1.43E-07	\\ 
 100,000	& 8.17E+12	& 8.17E+12	& -8.51E-08	\\ 
 1,000,000	& 6.01E+13	& 6.01E+13	& -3.46E-08	\\ \hline
\end{tabular}

\end{center}
\end{table}

\subsection{The case of $\beta > \gamma $}\label{ex_sec_case3}

Finally, we treat the case of Theorem \ref {th_main}(iv). We set $\sigma _S = 100$. 
Here $\mathrm {VaR}_\alpha (L)$ is too small compared with $\mathrm {VaR}_\alpha (L+S)$, 
so we compare the values of $\mathrm {VaR}_\alpha (L+S)$ and 
\begin{eqnarray}\label{approx_case4}
\mathrm {Approx} = \mathrm {VaR}_\alpha (S) + 
\frac{1}{k\gamma }\mathrm {VaR}_\alpha (S)^{\gamma + 1 - \beta }. 
\end{eqnarray}
The results are shown in Table \ref {table_case3}. 
We see that the error ($ = \mathrm {Approx} / \mathrm {VaR}_\alpha (L+S) - 1$) tends to become smaller when $\xi _S$ is large. 

Table \ref {table_case3} also indicates that 
the supplement of $S$ has quite a large effect on the risk amount when the distribution of $S$ has a fat tail. 
For example, when $\xi _S = 3.0$, the value of $\mathrm {VaR}_\alpha (L+S)$ is more than 
$90$ times $\mathrm {VaR}_\alpha (L)$ and is heavily influenced by the tail of $S$. 
We see that a small change of $\xi _S$ may cause a huge impact on the risk model. 

In our examples, we do not treat the case of Theorem \ref {th_main}(v), 
but a similar implication is obtained  in that case, too.

\begin{table}[!h]
\begin{center}
\caption{
Approximation errors in the case $\beta > \gamma $. 
$\xi _S$ is the shape parameter of the GPD with respect to $S$. 
Approx is given by (\ref {approx_case4}). 
Error $= \mathrm {Approx} / \mathrm {VaR}_\alpha (L+S) - 1$. 
}
\label{table_case3}
\begin{tabular}{@{}ccccc@{}} \hline
 $\xi _S$	& $\mathrm {VaR}_\alpha (L+S)$	& Approx	& Error	& $\mathrm {VaR}_\alpha (S)$	\\ \hline
 2.5 	& 2.12E+12	& 1.52E+12	& -2.82E-01	& 4.00E+11	\\ 
 3.0 	& 4.64E+13	& 4.56E+13	& -1.61E-02	& 3.34E+13	\\ 
 3.5 	& 2.99E+15	& 2.99E+15	& -3.04E-04	& 2.86E+15	\\ 
 4.0 	& 2.52E+17	& 2.52E+17	& -5.38E-06	& 2.50E+17	\\ 
 4.5 	& 2.23E+19	& 2.23E+19	& -2.09E-07	& 2.22E+19	\\ \hline
\end{tabular}
\end{center}
\end{table}

\section{Concluding Remarks}\label{sec_conclusion}

In this paper, we have introduced a theoretical framework of sensitivity analysis for quantitative operational risk management. 
Concretely speaking, we have investigated the impact on the risk amount (here, VaR) arising from adding the loss amount variable $S$ 
to the present loss amount variable $L$ 
when the tail probabilities of $L$ and $S$ are regularly varying 
($L \in \mathcal {R}_{-\beta }, S \in \mathcal {R}_{-\gamma }$ for some $\beta , \gamma  > 0$). 
The result depends on the relative magnitudes of $\beta $ and $\gamma $. 
One implication is that we must pay more attention to the form of the tail of $S$ when its tail is fatter. 
Nonetheless, as long as $\gamma > \beta + 1$, the difference between the prior VaR and the posterior VaR 
is approximated by the EL of $S$. 
As mentioned in the end of Section \ref {sec_intro}, 
similar phenomena were empirically found in \cite {Frachot-et-al}. 
As far as we know, this paper is the first to provide a theoretical rationale for these 
while paying attention to the difference between the tail indices of loss distributions. 

We have mainly treated the case where $L$ and $S$ are independent, 
except for a few cases discussed in Section \ref {sec_generalization}. 
In related studies of the case where $L$ and $S$ are dependent, 
\cite {Bocker-Kluppelberg2} 
invoke a L\'evy copula to describe the dependency and 
give an asymptotic estimate of Fr\'echet bounds of total VaR. 
In 
\cite {Embrechts-et-al3}, 
the sub- and super-additivity of quantile-based risk measures is 
discussed when $L$ and $S$ are correlated (see also Remark \ref {rem_dependency}). 
However, the focus of these studies is concentrated on the case of $\beta = \gamma $. 
Directions of our future work are to deepen our study and extend it to more general cases when $L$ and $S$ have a dependency structure 
(including the case of $\beta \neq \gamma $).

Another interesting topic is to study a similar asymptotic property of CVaRs, 
\begin{eqnarray}
\mathrm {CVaR}_\alpha (X) = \E [X |  X \geq \mathrm {VaR}_\alpha (X)]. 
\end{eqnarray}
Since CVaR is coherent and continuously differentiable (in the sense of \cite {Tasche2}), the following inequality holds in general (assuming $\E [L] + \E [S] < \infty $) 
\begin{eqnarray}
\Delta \mathrm {CVaR}^{L, S}_\alpha := 
\mathrm {CVaR}_\alpha (L+S) - \mathrm {CVaR}_\alpha (L) \leq 
\E [S | L+S \geq \mathrm {VaR}_\alpha (L+S)]  \ \ \ \ \ 
\end{eqnarray}
(see Proposition 17.2 in \cite {Tasche2}\footnote{The author thanks one of the anonymous referees for highlighting this point.}). 
Another remaining task is to investigate the property of $\Delta \mathrm {CVaR}^{L, S}_\alpha $ more precisely
when $\alpha $ is close to one. 

\def\thesection{\Alph{section}}
\setcounter{section}{0}
\section{Appendix}\label{sec_appendix}

\subsection{Consideration of dependency structure}\label{sec_generalization}

In Sections \ref {sec_sum_VaR}--\ref {sec_ex}, we have mainly assumed that $L$ and $S$ are independent, 
since they were caused by different loss factors. 
However, huge losses often happen due to multiple simultaneous loss events. 
Thus, it is important to allocate risk capital considering a dependency structure between loss factors. 
Basel II states that 
``scenario analysis should be used to assess the impact of deviations from the correlation assumptions 
embedded in the bank's operational risk measurement framework, in particular, to evaluate 
potential losses arising from multiple simultaneous operational risk loss events'' 
in paragraph 675 of 
\cite {BaselII}. 

In this section, we consider the case where $L$ and $S$ are not necessarily independent, and 
present generalizations of Theorem \ref {th_main}. 
Recall that $L\in \mathcal {R}_{-\beta }$ and $S\in \mathcal {R}_{-\gamma }$ are random variables 
for some $\beta , \gamma > 0$. 
We focus on the case of $\beta \leq  \gamma $. 
Let $p$ (resp. $q$) $: [0, \infty )\times \mathcal {B}([0, \infty )) \longrightarrow [0, 1]$ 
be a regular conditional probability distribution 
with respect to $F_L$ (resp., $F_S$) given $S$ (resp., $L$); that is, $P(L\in \cdot  | S) = p(S, \cdot )$ and 
$P(S\in \cdot  | L) = q(L, \cdot )$. 
Here, $\mathcal {B}([0, \infty ))$ is the Borel field of $[0, \infty )$. 
We define the function $F_L(x | S = s)$ by $F_L(x | S = s) = p(s , [0, x])$. 
We see that the function $F_L(x | S = s)$ satisfies 
\begin{eqnarray}
\int _BF_L(x | S = s)F_S(ds) = P(L \leq x, S\in B)
\end{eqnarray}
for each $B\in \mathcal {B}([0, \infty ))$. 

We prepare the following conditions. 
\begin{description}
 \item[ \mbox{[A5]}] \ There is some $x_0 > 0$ such that 
$F_L(\cdot | S = s)$ has a positive, non-increasing, continuous density function $f_L(\cdot | S = s)$ on $[x_0, \infty )$ 
for $P(S\in \cdot )$-a.a. $s$. 
 \item[ \mbox{[A6]}] \ It holds that 
\begin{eqnarray}\label{cond_D}
\mathop {{\rm ess}\sup}_{s\geq 0}\sup _{t\in K}\left| 
\frac{f_L(tx | S = s)}{f_L(x | S = s)} - t^{-\beta - 1}\right| \ \longrightarrow \ 0, \ \ 
x\rightarrow \infty 
\end{eqnarray}
for any compact set $K\subset (0, 1]$ and 
\begin{eqnarray}\label{cond_D2}
\int _{[0, \infty )}s^\eta \frac{f_L(x | S = s)}{f_L(x)}F_S(ds) \leq C, \ \ x \geq  x_0
\end{eqnarray}
for some constants $C > 0$ and $\eta > \gamma  - \beta $, 
where $\mathop {{\rm ess}\sup}$ is 
the $L^\infty $-norm under the measure $P(S\in \cdot )$. 
\end{description}

Let $\E [\cdot | L = x]$ be the expectation under the probability measure $q(x, \cdot )$. 
Under condition [A5], we see that for each $\varphi \in L^1([0, \infty ); P(S\in \cdot ))$ 
\begin{eqnarray}\label{cond_exp_L}
\E [\varphi (S) | L = x] = \int _{[0, \infty )}\varphi (s)\frac{f_L(x | S = s)}{f_L(x)}F_S(ds), \ \ 
P(L\in \cdot )\mbox {-a.a. } x\geq x_0. \ \ \ \ \ 
\end{eqnarray}
We do not distinguish the left- and right-hand sides of (\ref {cond_exp_L}). 
In particular, the left-hand side of (\ref {cond_D2}) is regarded as $\E [S^\eta | L = x]$. 

\begin{remark}\label{rem_A1} \ 
\begin{itemize}
 \item [(i)] Note that condition [A5] includes condition [A2]. 
Under [A5]--[A6], we have 
$P(L > x, S > x) \leq Cx^{-\eta }\bar{F}_L(x)$ 
and then the negligible joint tail condition [A1] is also satisfied 
(therefore, the dependency of $L$ and $S$ considered in this section is not so strong). 
 \item [(ii)] Conditions [A5] and [A6] seem to be a little strong, but we have an example. 
Let $U\in \mathcal {R}_{-\beta }$ be a non-negative random variable that is independent of $S$ and 
let $g(s)$ be a positive measurable function. We define 
\begin{eqnarray}\label{eg_EF}
L = g(S)U. 
\end{eqnarray}
If we assume that $a\leq g(s)\leq b$ for some $a, b > 0$ and 
that $F_U$ has a positive, non-increasing, continuous density function $f_U$, 
then we have $f_L(x | S = s) = f_U(x / g(s)) / g(s)$ and 
\begin{eqnarray}
 \frac{f_L(tx | S = s)}{f_L(x | S = s)} - t^{-\beta - 1} = 
\frac{f_U(tx/g(s))}{f_U(x/g(s))} - t^{-\beta - 1}. 
\end{eqnarray}
Since $g(s)$ has an upper bound, we see that $f_L(x | S = s)$ satisfies (\ref {cond_D}) 
by using Theorem 1.5.2 of 
\cite {Bingham-et-al}. 
Moreover, it follows that for $\eta \in (\gamma - \beta , \gamma )$ 
\begin{eqnarray}\label{temp_eg}
\E [S^\eta  | L = x] \leq \frac{b}{a}\E [S^\eta ]\frac{f_U(x/b)}{f_U(x/a)}, \ \ 
P(L\in \cdot )\mbox {-a.a. } x \geq  x_0
\end{eqnarray}
and the right-hand side of (\ref {temp_eg}) converges to 
$\left (b/a\right )^{\beta + 2}\E [S^\eta ]$ as $x\rightarrow \infty $. 
Thus (\ref {cond_D2}) is also satisfied. 

At this stage, we cannot find financial applications of (\ref {eg_EF}). 
One of our aims for the future is to give a more natural example satisfying [A5]--[A6], or 
to weaken the assumptions more than [A5]--[A6]. 
\end{itemize}
\end{remark}

Now we present the following theorem. 

\begin{theorem} \ \label{th_main2}
\begin{description}
 \item[ \mbox{(i)} ] \ Assume $\mathrm {[A5]}$ and $\mathrm {[A6]}$. If $\beta + 1 < \gamma $, then 
\begin{eqnarray}\label{EL_Corr}
\Delta \mathrm {VaR}_\alpha ^{L, S} \sim 
\E [S | L = \mathrm {VaR}_\alpha (L)], \ \ 
\alpha \rightarrow 1. 
\end{eqnarray}
 \item[ \mbox{(ii)} ] \ Assume $\mathrm {[A3]}$, $\mathrm {[A5]}$ and $\mathrm {[A6]}$. 
Then the same assertions as Theorem \ref {th_main} $(ii)(iii)$ hold. 
\end{description}
\end{theorem}

Relation (\ref {EL_Corr}) gives an implication similar to (5.12) in 
\cite {Tasche}. 
The right-hand side of (\ref {EL_Corr}) has the same form as the so-called component VaR: 
\begin{eqnarray}\label{M-VaR}
\E [S  | L+S = \mathrm {VaR}_\alpha (L+S)] = 
\frac{\partial }{\partial u}\mathrm {VaR}_\alpha (L+uS)\Big |_{u = 1} 
\end{eqnarray}
under some suitable mathematical assumptions. 
In Section \ref {sec_EL2} we study the details. 
We can replace the right-hand side of (\ref {EL_Corr}) with (\ref {M-VaR}) by 
a few modifications of our assumptions: 
\begin{description}
 \item[ \mbox{[A5']}] \ The same condition as [A5] holds by replacing $L$ with $L+S$. 
 \item[ \mbox{[A6']}] \ Relations (\ref {cond_D}) and (\ref {cond_D2}) hold by replacing $L$ with $L+S$ 
and by setting $K = [a, \infty )$ for any $a > 0$. 
\end{description}
Indeed, our proof also works upon replacing $(L+S, L)$ with $(L, L+S)$. 

\begin{remark}\label{rem_dependency}
In this section, we treat only the case where the dependency of $L$ and $S$ is not very strong 
(see Remark \ref {rem_A1}(i)). 
Needless to say, it is meaningful to study a more highly correlated case. 
In such a case, 
the dependence structure between $L$ and $S$ is more effective 
for the asymptotic behavior of $\Delta \mathrm {VaR}_\alpha ^{L, S}$. 
Therefore, it is not so easy to introduce a result like Theorem \ref {th_main} 
without deeply investigating the dependency of $L$ and $S$. 

When $\beta = \gamma $, 
the asymptotic behavior of $\mathrm {VaR}_\alpha (L+S)$ as $\alpha \rightarrow 1$ 
is studied in several papers within the framework of multivariate extreme value theory, 
including the case where $L$ and $S$ are highly correlated. 
For instance, the arguments in Section 6 in Embrechts et al.~(2009) show that 
when $L$ and $S$ are identically distributed (and thus $\beta = \gamma $) 
and the dependence of $L$ and $S$ is described by the Fr\'echet copula 
(say, $C^F_{L, S}(u, v)$), 
the asymptotic sub- and super-additivity of VaR for $L$ and $S$ is determined by 
the tail dependence coefficients of $C^F_{L, S}(u, v)$. 
\end{remark}

\subsection{Effect of a supplement of a small loss amount}\label{sec_EL2}

In this section, we treat modified versions of Theorems \ref {th_main}(i) and \ref {th_main2}(i). 
We do not assume that the random variables are regularly varying, but 
that the additional loss amount variable is very small. 
Let $L, \tilde{S}$ be non-negative random variables and let $\varepsilon > 0$. 
We define a random variable $S_\varepsilon $ by $S_\varepsilon  = \varepsilon \tilde{S}$. 
We regard $L$ (resp. $L + S_\varepsilon $) as 
the prior (resp. posterior) loss amount variable and 
consider the limit of the difference between the prior and posterior VaR 
when taking $\mathrm {\varepsilon }\rightarrow 0$. 
Instead of making assumptions of regular variation, we adopt ``Assumption $(S)$'' in 
\cite {Tasche}. 
Then Lemma 5.3 and Remark 5.4 in 
\cite {Tasche} 
imply 
\begin{eqnarray}\label{rel_Tasche}\nonumber 
\lim _{\varepsilon \rightarrow 0}
\frac{\mathrm {VaR}_\alpha (L + S_\varepsilon ) - \mathrm {VaR}_\alpha (L)}{\varepsilon } = 
\frac{\partial }{\partial \varepsilon }\mathrm {VaR}_\alpha (L + \varepsilon \tilde{S})\Big |_{\varepsilon = 0} = 
\E [\tilde{S} | L = \mathrm {VaR}_\alpha (L)]. \\
\end{eqnarray}

By (\ref {rel_Tasche}), we have 
\begin{eqnarray}\nonumber 
\mathrm {VaR}_\alpha (L + S) - \mathrm {VaR}_\alpha (L) 
&=& 
\E [S | L = \mathrm {VaR}_\alpha (L) ] + o(\varepsilon ) \\
(  &=& 
\E [S | L+S_0 = \mathrm {VaR}_\alpha (L+S_0) ] + o(\varepsilon ) \ ) , 
\ \ \varepsilon \rightarrow 0, \ \ \ \ \ \ \ \ \ \ \ 
\label{rel_EL2}
\end{eqnarray}
where we simply put $S = S_\varepsilon $. 
In particular, if $L$ and $S$ are independent, then
\begin{eqnarray}
\mathrm {VaR}_\alpha (L + S) - \mathrm {VaR}_\alpha (L) \ = \ 
\E [S] + o(\varepsilon ) , \ \ \varepsilon \rightarrow 0. 
\end{eqnarray}
Thus, the effect of a supplement of the additional loss amount variable $S$ is 
approximated by its component VaR or EL. 
So, assertions similar to Theorems \ref {th_main}(i) and \ref {th_main2}(i) also hold in this case. 

The concept of the component VaR is related to the theory of risk capital decomposition 
(or risk capital allocation). 
Let us consider the case where $L$ and $S$ are loss amount variables and 
the total loss amount variable is given by $T(w_1, w_2) = w_1L + w_2S$ 
with a portfolio $(w_1, w_2)\in \Bbb {R}^2$. 
We try to calculate the risk contributions for the total risk capital $\rho (T(w_1, w_2))$, 
where $\rho $ is a risk measure. 

One idea is to apply Euler's relation 
\begin{eqnarray}
\rho (T(w_1, w_2)) = w_1\frac{\partial }{\partial w_1}\rho (T(w_1, w_2)) + 
w_2\frac{\partial }{\partial w_2}\rho (T(w_1, w_2))
\end{eqnarray}
when $\rho $ is linear homogeneous and $\rho (T(w_1, w_2))$ is differentiable with respect to $w_1$ and $w_2$. 
In particular we have 
\begin{eqnarray}\label{rel_Euler}
\rho (L + S) = \frac{\partial }{\partial u}\rho (uL + S)\Big |_{u=1} + 
\frac{\partial }{\partial u}\rho (L + uS)\Big |_{u=1}
\end{eqnarray}
and the second term in the right-hand side of (\ref {rel_Euler}) is regarded as the risk contribution of $S$. 
As in early studies considering the case of $\rho = \mathrm {VaR}_\alpha $, 
the same decomposition as (\ref {rel_Euler}) is obtained in 
\cite {Garman} 
and 
\cite {Hallerbach} 
and the risk contribution of $S$ is called the component VaR. 
The consistency of the decomposition of (\ref {rel_Euler}) has been studied from several points of view
(
\cite {Denault}, 
\cite {Kalkbrener}, 
\cite {Tasche},
and so on). 
In particular, Theorem 4.4 in 
\cite {Tasche} 
implies that the decomposition of (\ref {rel_Euler}) is 
``suitable for performance measurement'' (Definition 4.2 of 
\cite {Tasche}). 
Although many studies assume that $\rho $ is a coherent risk measure, 
the result of 
\cite {Tasche} 
also applies to the case of $\rho = \mathrm {VaR}_\alpha $. 

Another approach toward calculating the risk contribution of $S$ is 
to estimate the difference between the risk amounts $\rho (L+S) - \rho (L)$, 
which is called the marginal risk capital---see 
\cite {Merton-Perold}. 
(When $\rho = \mathrm {VaR}_\alpha $, it is called a marginal VaR.) 
This is intuitively understandable, however the aggregation of marginal risk capitals is not equal to 
the total risk amount $\rho (L+S)$. 

Relation (\ref {rel_EL2}) gives the equivalence between the marginal VaR and the component VaR 
when $S ( = \varepsilon \tilde{S})$ is very small. 
Theorem \ref {th_main2}(i) implies that 
the marginal VaR and the component VaR are also (asymptotically) equivalent 
when $L$ and $S$ have regularly varying tails and the tail of $S$ is sufficiently thinner than that  of $L$.

\subsection{Proofs}\label{sec_proofs}

In this section we present the proofs of our results. 
First, we list the auxiliary lemmas used to show our main results. 

\begin{lemma} \ \label{lemma1}
Let $X, Y$ be nonnegative random variables satisfying 
$\bar{F}_X\in \mathcal {R}_{-\beta }$ and $\bar{F}_Y\in \mathcal {R}_{-\gamma }$ 
for $\beta , \gamma > 0$. 
Assume the negligible joint tail condition [A1] (replacing $L$ and $S$ with $X$ and $Y$). 
Then $\bar{F}_{X + Y}(x) \sim \bar{F}_X(x) + \bar{F}_Y(x)$ as $x\rightarrow \infty $. 
Moreover $\bar{F}_{X+Y}\in \mathcal {R}_{-\min \{ \beta , \gamma \} }$. 
\end{lemma}

\begin{proof}
When $\beta < \gamma $, we see that 
\begin{eqnarray}
\liminf _{x\rightarrow \infty }\frac{\bar{F}_{X+Y}(x)}{\bar{F}_X(x) + \bar{F}_Y(x)}\geq 
\liminf _{x\rightarrow \infty }\frac{\bar{F}_X(x)}{\bar{F}_X(x) + \bar{F}_Y(x)} = 1. 
\end{eqnarray}
Similarly, for each $\varepsilon \in (0, 1)$, 
\begin{eqnarray}\label{temp_eq}
\limsup _{x\rightarrow \infty }\frac{\bar{F}_{X+Y}(x)}{\bar{F}_X(x) + \bar{F}_Y(x)}\leq 
\limsup _{x\rightarrow \infty }\frac{\bar{F}_X((1 - \varepsilon )x) + \bar{F}_Y(\varepsilon x)}{\bar{F}_X(x)} = 
(1 - \varepsilon )^{-\beta }. \ \ \ \ \ 
\end{eqnarray}
Since $\varepsilon $ is arbitrary, the left-hand side of (\ref {temp_eq}) is bounded from above by 1. 
Therefore, we observe $\bar{F}_{X+Y}\sim \bar{F}_X + \bar{F}_Y\sim \bar{F}_X\in \mathcal {R}_{-\beta } = \mathcal {R}_{-\min \{\beta , \gamma \}}$, and 
so the assertions are true. 

The above arguments also work well in the case of $\beta > \gamma $ by switching the roles of $X$ and $Y$. 
When $\beta = \gamma $, the assertions are given by Lemma 4 in 
\cite {Jang-Jho}. 
\end{proof}

\begin{remark}
In the above lemma, condition [A1] is required only when $\beta = \gamma $. 
The assertions always hold in general when $\beta \neq \gamma $. 
\end{remark}

The following Lemma \ref {lemma2} is strongly related to Theorem 2.4 in 
\cite {Bocker-Kluppelberg} 
and 
Theorem 2.14 in 
\cite {Bocker-Kluppelberg2} 
when $\beta = \gamma $. 

\begin{lemma} \ \label{lemma2}
Let $X\in \mathcal {R}_{-\beta }$, $Y\in \mathcal {R}_{-\gamma }$ be random variables with $\beta , \gamma > 0$. 
We assume that $\bar{F}_X(x)\sim \lambda \bar{F}_Y(x^{\beta / \gamma })$, 
$x\rightarrow \infty $ for some $\lambda  > 0$. 
Then $\mathrm {VaR}_\alpha (X)\sim \mathrm {VaR}_{1-(1-\alpha )/\lambda }(Y)^{\gamma / \beta }\sim 
\lambda ^{1/\beta }\mathrm {VaR}_\alpha (Y)^{\gamma / \beta }$, $\alpha \rightarrow 1$. 
\end{lemma}

\begin{proof} 
For $\xi \in (1, \infty )$, we put 
$v_X(\xi ) = \mathrm {VaR}_{1-1/\xi }(X)$ and 
$v_Y(\xi ) = \mathrm {VaR}_{1-1/\xi }(Y)$. 
Note that $v_X(\xi )$ (resp. $v_Y(\xi )$) 
is a left-continuous version of the generalized inverse function of $1/\bar{F}_X$ (resp. $1/\bar{F}_Y$) 
defined in 
\cite {Bocker-Kluppelberg}. 
By Proposition 2.13 in 
\cite {Bocker-Kluppelberg}, 
we have 
$v_X \in \mathcal {R}_{1/\beta }$ and $v_Y\in \mathcal {R}_{1/\gamma }$. 

By Theorem 1.5.12 in 
\cite {Bingham-et-al}, 
we get 
$(1/\bar{F}_X)(v_X(\xi )) \sim \xi $ and 
$(1/\bar{F}_Y)( v_Y(\lambda \xi ) ) \sim \lambda \xi $ as $\xi \rightarrow \infty $. 
Thus 
\begin{eqnarray}
\bar{F}_X(v_Y(\lambda \xi )^{\gamma / \beta }) \sim 
\lambda \bar{F}_Y(v_Y(\lambda \xi )) \sim \bar{F}_X(v_X(\xi )) \sim 
\lambda \bar{F}_Y(v_X(\xi )^{\beta / \gamma }), \ \ \xi \rightarrow \infty . \ \ \ \ \ \ \ \ 
\end{eqnarray}
Then we have $v_Y(\lambda \xi )\sim v_X(\xi )^{\beta / \gamma }$ and 
$v_X(\xi )\sim v_Y(\lambda \xi )^{\gamma / \beta }\sim \lambda ^{1/\beta }v_Y(\xi )^{\gamma / \beta }$ as 
$\xi \rightarrow \infty $, 
which imply our assertions. 
\end{proof}

The following lemma is easily obtained from Theorem A3.3 in 
\cite {Embrechts-et-al}. 

\begin{lemma} \ \label{lem_varying}
Let $f$ be a regularly varying function and let 
$(x_n)_n, (y_n)_n\subset (0, \infty )$ be such that $x_n, y_n \longrightarrow \infty $ and 
$x_n \sim y_n$ as $n\rightarrow \infty $. Then $f(x_n) \sim f(y_n)$. 
\end{lemma}

We are now ready to prove the main results. 

\noindent
{\it Proof of Proposition \ref {prop_sum_VaR}.} 
All assertions are obtained from Lemmas \ref {lemma1}--\ref {lemma2}. 
\hfill $\Box$

\noindent
{\it Proof of Theorem \ref {th_main2}$(i)$.} 
Define 
\begin{eqnarray}
l(x, s) = s\frac{f_L(x | S = s)}{f_L(x)}, \ \ 
K(x) = \int _{[0, \infty )}l(x, s)F_s(ds) = \E [S | L = x], \ \ x\geq x_0. \ \ \ \ \ \ \ \ 
\end{eqnarray}
Since $\eta > \gamma - \beta > 1$, the relation (\ref {cond_D2}) implies that 
$\left( l(x, \cdot )\right)_{x\geq x_0}$ is 
uniformly integrable with respect to $P(S\in \cdot )$. 
Thus, $K(x)$ is continuous in $x \geq x_0$. 
Moreover, since it follows that 
\begin{eqnarray}\nonumber 
&&|K(tx) - K(x)|\\\nonumber 
&\leq & 
\int _{[0, \infty )}s\cdot \frac{f_L(x | S = s)}{f_L(x)}
\left| \frac{f_L(tx | S = s)}{f_L(x | S = s)}\cdot \frac{f_L(x)}{f_L(tx)} - 1\right| F_S(ds)\\\nonumber 
&\leq &
\left\{  \mathop {{\rm ess}\sup}_{s\geq 0}\left| 
\frac{f_L(tx | S = s)}{f_L(x | S = s)} - t^{-\beta - 1}\right| + 
\left| 
\frac{f_L(x)}{f_L(tx)} - t^{\beta + 1}\right| \right\} \\
&&\times 
\left( \left| 
\frac{f_L(x)}{f_L(tx)} - t^{\beta + 1}\right| + 2t^{\beta + 1} \right) K(x)
\end{eqnarray}
for each $t > 0$, we see that $K\in \mathcal {R}_0$ by virtue of (\ref {cond_D}). 

We prove the following proposition. 

\begin{proposition} \ \label{prop_proof_th_EL}
$\frac{F_{L+S}(x) - F_L(x)}{f_L(x)} \ \sim \ - K(x), \ \ 
x\rightarrow \infty $. 
\end{proposition}

\begin{proof} 
By the assumptions $L, S \geq 0$ and [A5], we have 
\begin{eqnarray}\label{rel_FL1}
F_{L+S}(x) - F_L(x) \ = \ 
- I^1(x) + I^2(x) - I^3(x) 
\end{eqnarray}
for $x > 2x_0$, where 
\begin{eqnarray}
I^1(x) &=& \int ^1_0\int _{[0, x/2]}f_L(x - us | S = s)sF_S(ds)du, \\
I^2(x) &=& P( L + S \leq x, \ x/2 < S \leq x), \\
I^3(x) &=& P( L \leq x, \ S > x/2) . 
\end{eqnarray}
Since $f_L\in \mathcal {R}_{-\beta - 1}$, $\bar{F}_S\in \mathcal {R}_{-\gamma }$ and $K\in \mathcal {R}_0$, 
we have 
\begin{eqnarray}\label{res_lemma3_1}
\frac{I^2(x)}{f_L(x)K(x)} + \frac{I^3(x)}{f_L(x)K(x)} \ \leq \ 
\frac{2\bar{F}_S(x/2)}{f_L(x)K(x)} 
\ \longrightarrow \ 0, \ \ x\rightarrow \infty . 
\end{eqnarray}
To estimate the term $I^1(x)$, we define a random variable $T$ by $T = S/x$ and a function $J(x)$ by 
\begin{eqnarray}
J(x) = \int ^1_0\int _{[0, x/2]}(1-us/x)^{-\beta - 1}s
\frac{f_L(x | S = s)}{f_L(x)}F_S(ds)du. 
\end{eqnarray}
Then assumption [A6] implies 
\begin{eqnarray}\nonumber 
&&\frac{1}{K(x)}\left| I^1(x)/f_L(x) - J(x) \right| \\
&\leq & \nonumber 
\frac{1}{K(x)}\int ^1_0\int _{[0, 1/2]}xt
\left| \frac{f_L((1-ut)x | S = tx)}{f_L(x | S = tx)} - (1-ut)^{-\beta - 1}\right| 
\frac{f_L(x | S = tx)}{f_L(x)}F_T(dt)du\\\label{res_lemma3_3}
&\leq & 
\mathop {\rm {ess} \sup }_{s\geq 0}\sup _{r\in [1/2,1]}
\left| \frac{f_L(rx | S = s)}{f_L(x | S = s)} - r^{-\beta - 1}\right| 
\ \longrightarrow  \ 0, \ \ x \rightarrow \infty . 
\end{eqnarray}
Moreover we can rewrite $J(x)$ as 
\begin{eqnarray}
J(x) = 
\int _{[0, x/2]}\frac{\left( 1-sy\right) ^{-\beta } - 1}{\beta y}\cdot \frac{f_L(x | S = s)}{f_L(x)}F_S(ds), 
\end{eqnarray}
where $y = 1/x$. Then Taylor's theorem implies 
\begin{eqnarray}\nonumber 
&&|J(x) - K(x)| \\\nonumber 
&\leq & \int _{(x/2, \infty )}l(x, s)F_S(ds) + 
\int _{[0, x/2]}\frac{\left| \left( 1-sy\right) ^{-\beta } - 1 - \beta sy\right| }{\beta y} 
\cdot \frac{f_L(x | S = s)}{f_L(x)}F_S(ds)\\\nonumber 
&\leq & 
\frac{2^{\eta - 1}C}{x^{\eta - 1}} + (\beta + 1)y
\int _{[0, x/2]}s^2\int ^1_0(1-u)(1-usy)^{-\beta - 2}du\frac{f_L(x | S = s)}{f_L(x)}F_S(ds)\\
&\leq & 
\frac{2^{\eta - 1}C}{x^{\eta - 1}} + \frac{2^{\beta + \tilde{\eta }}(\beta + 1)C}{x^{\tilde{\eta }- 1}}, 
\end{eqnarray}
where $\tilde{\eta } = \min \{ \eta , 2 \}$. Thus 
\begin{eqnarray}\label{res_lemma3_4}
|J(x) / K(x) - 1|\ \longrightarrow \ 0, \ \ x\rightarrow \infty. 
\end{eqnarray}
By (\ref {res_lemma3_1})--(\ref {res_lemma3_4}), 
we obtain the assertion. 
\end{proof}

Now we complete the proof of Theorem \ref {th_main2}(i). 
Let us put $x_\alpha = \mathrm {VaR}_\alpha (L)$ and 
$y_\alpha = \mathrm {VaR}_\alpha (L + S)$. 
Obviously $y_\alpha \geq x_\alpha $ and 
we may assume $x_\alpha > x_0$ ($x_0$ is given in [A5]). 
Since $\alpha = F_L(x_\alpha ) = F_{L+S}(y_\alpha )$, we have 
\begin{eqnarray}\label{temp_proof_th_EL_0}
-\frac{F_{L+S}(y_\alpha ) - F_L(y_\alpha )}{f_L(y_\alpha )} = 
\frac{F_L(y_\alpha ) - F_L(x_\alpha )}{f_L(y_\alpha )} = 
\int ^1_0g_\alpha (u)du(y_\alpha - x_\alpha ), 
\end{eqnarray}
where $g_\alpha (u) = f_L(x_\alpha + u(y_\alpha - x_\alpha ))/f_L(y_\alpha )$. 
Proposition \ref {prop_proof_th_EL} implies that the left-hand side of (\ref {temp_proof_th_EL_0}) 
is asymptotically equivalent to $K(y_\alpha )$. 
Moreover, using Proposition \ref {prop_sum_VaR}(i) and Lemma \ref {lem_varying}, 
we have $K(y_\alpha ) \sim K(x_\alpha )$ as $\alpha \rightarrow 1$. 
On the other hand, 
\begin{eqnarray}\label{temp_proof_th_EL_1}
1 \ \leq \ \int ^1_0g_\alpha (u)du \ \leq \ \frac{f_L(x_\alpha )}{f_L(y_\alpha )}
\end{eqnarray}
and so the right-hand side of (\ref {temp_proof_th_EL_1}) converges to one as $\alpha \rightarrow 1$
by using Proposition \ref {prop_sum_VaR}(i) and Lemma \ref {lem_varying} again. 
Thus, the right-hand side of (\ref {temp_proof_th_EL_0}) 
is asymptotically equivalent to $y_\alpha - x_\alpha $. Then we obtain the assertion. 
{\hfill $\square$}

\noindent 
{\it Proof of Theorem \ref {th_main2}$(ii)$.} 
When $\beta = \gamma $, the assertion is obtained from Remark \ref {rem_A1}(i) and 
Lemmas \ref {lemma1}--\ref {lemma2}, 
so we consider only the case of $\beta < \gamma \ (\leq \beta + 1)$. 
The assertion is obtained by a proof similar to that of 
Theorem \ref {th_main2}(i) by using the following proposition instead of Proposition \ref {prop_proof_th_EL}. 
{\hfill $\square$}

\begin{proposition} \ \label{prop_proof_th_thin}It holds that 
\begin{eqnarray}
\frac{F_{L+S}(x) - F_L(x)}{f_L(x)} \ \sim \ -\frac{kx^{\beta + 1 - \gamma }}{\beta }, \ \ 
x\rightarrow \infty . 
\end{eqnarray}
\end{proposition}

\begin{proof} 
Take any $0 < \varepsilon  < 1$. 
The same calculation as in the proof of Proposition \ref {prop_proof_th_EL} gives us 
\begin{eqnarray}
F_{L+S}(x) - F_L(x) \ = \ 
- I^1_\varepsilon (x) + I^2_\varepsilon (x) - I^3_\varepsilon (x) , 
\end{eqnarray}
where $I^j_\varepsilon (x)$ is the same as $I^j(x)$ on replacing 
$x/2$ with $(1 - \varepsilon )x$ ($j = 1, 2, 3$.) 
By assumption [A3] and the monotone density theorem, we see that 
\begin{eqnarray}\nonumber 
&&\limsup _{x\rightarrow \infty }\frac{I^2_\varepsilon (x)}{x^{\beta + 1 - \gamma }f_L(x)}
\ \leq \ 
\lim _{x\rightarrow \infty }\frac{\bar{F}_S((1-\varepsilon )x) - \bar{F}_S(x)}{x^{\beta + 1 - \gamma }f_L(x)}\\\label{temp_proof_th_thin1}
&&=\ 
\lim _{x\rightarrow \infty }\frac{\bar{F}_L(x)}{xf_L(x)}\cdot \frac{x^{\gamma - \beta }\bar{F}_S(x)}{\bar{F}_L(x)}\cdot 
\left( \frac{\bar{F}_S((1-\varepsilon )x)}{\bar{F}_S(x)} - 1 \right) \ =\ 
\frac{k}{\beta }((1-\varepsilon )^{-\gamma } - 1). \ \ \ \ \ \ \ \ \ 
\end{eqnarray}

By a calculation similar to the one in the proof of Proposition \ref {prop_proof_th_EL}, we get 
\begin{eqnarray}\label{temp_proof_th_thin2}
\frac{I^1_\varepsilon (x)}{x^{\beta + 1 - \gamma }f_L(x)}
\ \leq \ \frac{C'_\varepsilon }{x^{\beta + 1 - \gamma }}\int _{[0, x]}l(x, s)F_S(ds)
\end{eqnarray}
for some positive constant $C'_\varepsilon $. 
Assumption [A6] implies that the right-hand side of (\ref {temp_proof_th_thin2}) converges to zero 
as $x\rightarrow \infty $ for each $\varepsilon $. 
Indeed, if $\eta \geq  1$ then this is obvious. 
If $\eta < 1$, we have 
\begin{eqnarray}\nonumber 
&&\frac{1}{x^{\beta + 1 - \gamma }}\int _{[0, x]}l(x, s)F_S(ds)\\
&\leq & 
\frac{1}{x^{\beta - \gamma + \eta }}\int _{[0, x]}s^\eta \frac{f_L(x | S = s)}{f_L(x)}F_S(ds) \ \leq \ 
\frac{C}{x^{\beta - \gamma + \eta }}\ \longrightarrow \ 0, \ \ 
x\rightarrow \infty . \ \ \ \ \ 
\end{eqnarray}
Thus we get 
\begin{eqnarray}\label{temp_proof_th_thin3}
\lim _{x\rightarrow \infty }\frac{I^1_\varepsilon (x)}{x^{\beta + 1 - \gamma }f_L(x)} \ = \ 0. 
\end{eqnarray}

By assumption [A6], we have 
\begin{eqnarray}\nonumber 
&&\frac{1}{\bar{F}_S(x)}\int _{((1-\varepsilon )x, \infty )}\bar{F}_L(x | S = s)F_S(ds) = 
\frac{1}{\bar{F}_S(x)}\int ^\infty _xq(y, \{ S > (1-\varepsilon )x \})F_L(dy)\\
&&\leq 
\frac{1}{(1-\varepsilon )^{\eta }x^\eta \bar{F}_S(x)}\int ^\infty _x\E [S^\eta | L = y]F_L(dy)
\leq 
\frac{C\bar{F}_L(x)}{(1-\varepsilon )^{\eta }x^\eta \bar{F}_S(x)}, 
\end{eqnarray}
where $\bar{F}_L(x | S = s) = 1 - F_L(x | S = s)$. 
Then it holds that 
\begin{eqnarray}\label{temp_proof_th_thin4}\nonumber 
&&\frac{I^3_\varepsilon (x)}{x^{\beta + 1 - \gamma }f_L(x)} = 
\frac{x^{\gamma - \beta }\bar{F}_S(x)}{\bar{F}_L(x)}\\\nonumber 
&&\hspace{8mm}\times \frac{\bar{F}_L(x)}{xf_L(x)}
\left\{ 
\frac{\bar{F}_S((1-\varepsilon )x)}{\bar{F}_S(x)} - 
\frac{1}{\bar{F}_S(x)}\int _{((1-\varepsilon )x, \infty )}\bar{F}_L(x | S = s)F_S(ds)\right\} \\
&&\longrightarrow 
\frac{k(1-\varepsilon )^{-\gamma }}{\beta }, \ \ x\rightarrow \infty 
\end{eqnarray}
by virtue of the monotone density theorem and assumption [A3]. 

The relations (\ref {temp_proof_th_thin1}), (\ref {temp_proof_th_thin3}), (\ref {temp_proof_th_thin4}) and 
$I^2_\varepsilon \geq 0$ give us 
\begin{eqnarray}\nonumber 
-\frac{k(1-\varepsilon )^{-\gamma }}{\beta } = 
\liminf _{x\rightarrow \infty }\frac{F_{L+S}(x) - F_L(x)}{x^{\beta + 1 - \gamma }f_L(x)} \leq  
\limsup _{x\rightarrow \infty }\frac{F_{L+S}(x) - F_L(x)}{x^{\beta + 1 - \gamma }f_L(x)}
\leq -\frac{k}{\beta }. \\
\end{eqnarray}
Then we obtain the assertion by letting $\varepsilon \rightarrow 0$. 
\end{proof}

\noindent 
{\it Proof of Theorem \ref {th_main}. }
Assertions (i)--(ii) are obtained by the same (or easier) arguments as the proof of Theorem \ref {th_main2}. 
Assertion (iii) is a consequence of Lemmas \ref {lemma1}--\ref {lemma2}. 
Assertions (iv)--(v) follow from assertions (i)--(ii) by replacing the roles of $L$ and $S$. 
\hfill $\Box $

\section*{Acknowledgments}

The author would like to thank the anonymous referees for their valuable comments and suggestions, which have improved the quality of the paper.

\end{document}